\documentclass[12pt]{article}

\newcommand{\beq}{\begin{equation}}
\newcommand{\eeq}{\end{equation}}
\newcommand{\beqa}{\begin{eqnarray}}
\newcommand{\eeqa}{\end{eqnarray}}
\newcommand{\beqar}{\begin{eqnarray*}}
\newcommand{\eeqar}{\end{eqnarray*}}

\def \la {\langle}
\def \ra {\rangle}

\begin{document}

\input epsf
\title{\bf  \large
    Remote operations and interactions for \\
  systems of arbitrary dimensional Hilbert space: \\
          a state-operator  approach}

\author{
 Benni Reznik$^{a}$, Yakir Aharonov$^{a,b}$ and Berry Groisman$^{a}$
 {\ }\\
({\small \it a}) {\em \small School of Physics and Astronomy, Tel Aviv
University, Tel Aviv 69978, Israel}\\ ({\small\it b}) {\em \small Department
of Physics, University of South Carolina, Columbia, SC 29208}
}

\maketitle

\begin{abstract}
We present a systematic simple method for constructing
deterministic remote operations on single and multiple systems
of arbitrary discrete dimensionality.
These operations include remote rotations, remote interactions and
 measurements.
The resources needed for an operation on a two-level system
are one ebit and a bidirectional communication of two cbits, and
for an n-level system, a pair of entangled n-level particles
and two classical ``nits''.  In the latter case, there are $n-1$
possible distinct operations per one n-level entangled pair.
Similar results apply for generating interaction between a pair of
remote systems and for remote measurements.
We further consider remote operations on $N$
spatially distributed systems, and show that the number of
possible distinct operations increases here exponentially,
with the available number of entangled pairs that are initially
distributed between the systems.
Our results follow from the properties of a hybrid state-operator
object (``stator''),
which describes quantum correlations between states and operations.

\end{abstract}

\section{Introduction}

Over the recent years entanglement has been examined
as a resource which allows new types of communication tasks
such as teleportation,  dense coding,
and other local manipulations of entanglement
\cite{werner}.
These studies exploit the relation between quantum non-locality
and the structure of the Hilbert space.
A more recent avenue of research  examines the relation between
entanglement and the dynamical evolution of several systems.
Here, two basic questions have been examined: First, what is
the entanglement creation capability of a given Hamiltonian that
acts on a pair of systems \cite{zalka}.
The second question deals with the reverse problem:
 what types of non-local operations on two or
more remote systems can be generated, using a given resource
of entangled states, by applying
local operations and performing classical communication (LOCC).

In this article we will be interested in the second question.
Previous work has demonstrated that
certain operations like a remote  controlled-not (CNOT),
may consume less entanglement than what is needed when applying
teleportation techniques\cite{resources}.
For probabilistic non-local operations,
an isomorphism between the physical operations and the required
entanglement has been discovered, which for
certain operations necessitates less than one ebit per operation\cite{cirac}.
A closely related question, raised by Huelga {\it et. al.}\cite{huelga},
concerns the possibility of implementing a unitary
transformation on a remote system.

The purpose of this article is to  present a systematic
approach for constructing  a class of
deterministic remote unitary transformation,
and remote interactions between several distributed systems.
We assume that the parties share entangled states and are allowed
to perform only local operations and (bidirectional) classical communication.

A special characteristic of our method is that the
generators which give rise to the transformation,
are controlled locally by the two  parties.
The structure of the complete operation is in a sense
``split'' and determined by the local observers that posses the
distributed parts of the system.
Therefore, in the special case that the generators are known only locally,
one cannot perform the operation using ordinary teleportation
techniques.

To clarify this, consider the remote unitary operation
\beq
U_{B} = \exp\biggl[ i\alpha \sigma_{n_B}\biggr]
\eeq
that Alice and Bob wish to apply on a state $|\Psi_B\ra$ of Bob.
The axis $n_B$,  which defines $\sigma_{n_B}= \bar n_B\cdot
\bar \sigma_B$, is determined by Bob, while the angle of
rotation, $\alpha$, by Alice.

Similarly, if Alice and Bob wish to apply a remote
interaction
\beq
U_{AB} = \exp\biggl[ i\alpha \sigma_{n_A}\sigma_{n_B}\biggr]
\eeq
on a pair of spins in some arbitrary state $|\Psi_{AB}\ra$,
with one spin at the hands of Alice and the other with Bob,
then, the axes $n_A$ and $n_B$, which fix the local generators $\sigma_{n_A}$
and $\sigma_{n_B}$ are  controlled locally
by Alice and Bob respectively.

Our approach relies on the properties of a new hybrid object
which we introduce in section 2. This  object describes quantum
correlations between states of one party, say Alice, and operations acting
on an arbitrary state of Bob.
It turns our that   certain
remote operations can be translated to certain properties
of this hybrid state-operator object, which we will refer to
as a ``stator''. The possible remote operations are hence associated with properties
of the stator alone and are independent of the nature of the state(s)
upon we intend to act remotely.
By identifying the appropriate stator we are
able to apply a remote operation on an arbitrary state(s).

In section 3. we describe the physical context in which stators
can be prepared by applying LOCC on shared entanglement and the system.
In section 4. we  show how to use stators
to construct remote rotations for a 2-level (spin-half) system.
Then, in section 5. we consider the general problem of operating on an n-level system.
In section 6. we study the case of $N$ multiple systems,
and in section 7. we show how to promote remote unitary operations
into remote interactions and measurements.

\section{The Stator}

We begin by introducing a new object, which we shall
refer to as a ``stator''. A stator is
a hybrid linear construction of states in Alice's Hilbert space and
operators acting on Bob's system.
The purpose of introducing this object is twofold:
First, stators simplify considerably
the construction of remote unitary operations and interactions
via entanglement by providing us with a systematic general approach
which can be easily generalized to an arbitrary number of n-level system.
Second, we found that these objects, which describe
quantum correlations between states on one side and operators on the
other side, assist us to  develop an intuition  regarding
remote operations which may turn out helpful in other problems.

Let us then begin by defining what is a stator.
We denote the Hilbert spaces of two remote observers, Alice and
Bob, by $\cal{H}_A$ and $\cal{H}_B$, respectively.  Instead of
describing quantum correlations between states of Alice and Bob, we wish
now to
describe quantum correlations between {\em states} in $\cal{H}_A$ and resulting
{\em actions} described by
operators, in $ O(\cal{H}_B)$, acting on an {\em arbitrary} state in $\cal{H}_B$.
Hence we now construct a hybrid state-operator or shortly a ``stator'',
$\cal{S}$, that lives in the space

\beq
{\cal{S}} \in \{ \cal{H}_A\times O(\cal{H}_B) \}
\eeq
In close analogy to an entangled state,  a stator has the general form
\beq
{\cal{S}}= \sum_{ij}c_{ij}|i_A\ra\otimes  O_{Bj}
\eeq
with $|i_A\ra\in {\cal {H}}_A$, $O_{Bj}\in O({\cal{H}}_B)$,
and $c_{ij}$ as c-numbers.

Although this structure resembles the form of an entangled state
it does not describe a fixed amount of entanglement because it
applies to any general state of Bob. When we act
with a stator on a general state $|\Psi_B\ra\in {\cal H}_B$ we get
\beq
{\cal S} |\Psi_B\ra \in {\cal H}_A\otimes{\cal H}_B
\label{ABstate}
\eeq
Therefore, even if the stator has a maximal entanglement-like structure
(as in eq. (\ref{maximal}) below),
 the measure  of entanglement, $E({\cal S} |\Psi_B\ra)$,
depends on the nature of $|\Psi_B\ra$. For a general state of Bob,
we may get any value between of  $E({\cal S} |\Psi_B\ra)$,
from  zero to one even for a ``maximal'' stator.

Most important to us will be the following property.
For every stator we can construct an {\em eigenoperator equation}
\beq
O_A {\cal{S}} = \lambda_B {\cal{S}}
\eeq
Thus,  by operating on the stator with an operator $O_A\in{\cal {H}}_A$
in Alice's Hilbert space, we get back the same stator multiplied by
an {\em eigenoperator} now  acting in Bob's Hilbert space $\cal {H_B}$.

In general, the operators $O_A$ and eigenoperators $\lambda_B$
need not be hermitian. However, as we shall see, for
certain classes of stators, relevant to the present problem, the operators and
eigenoperators are both hermitian.

As a first example, let ${\rm dim}{\cal H}_A=2$ be spanned by the
eigenstates  $|0_A\ra$ and $|1_A\ra$ of $\sigma_{z_A}$,
and consider the operator $\sigma_{n_B}\in O(\cal{H}_B)$
such that $\sigma_{n_B}^2 = I_{B}$.
Consider now the stator
\beq
{\cal{S}}= |0_A\ra \otimes I_{B} + |1_A\ra \otimes \sigma_{n_B}
\label{maximal}
\eeq
which we shall refer to in the sequal as a 2-level stator.
${\cal S}$ satisfies the eigenoperator equation:
\beq
\sigma_{x_A} {\cal{S}} = \sigma_{n_B} {\cal{S}}
\eeq

As straightforward, but useful consequence,
any analytic function  $f$  also satisfies
\beq
f( \sigma_{x_A} ) {\cal{S}} = f (\sigma_{n_B}) {\cal{S}}
\eeq
and particularly
\beq
e^{i\alpha \sigma_{x_A} } {\cal{S}} = e^{i\alpha \sigma_{n_B}} {\cal{S}}
\eeq
where $\alpha$ is any real number or hermitian operator in ${\cal H}_A$.
The relation above
already indicates why stators can be useful  for generating remote operations.
We note that a unitary operation of Alice gives rise to a similar
unitary operation acting on Bob's side.

The above construction can be generalized to the  case
${\rm dim}{\cal {H}_A} = n$, which becomes relevant if Bob owns
an n-level system.
Let $|i_B \ra$, $i=0,1\cdots n-1$, be an orthogonal basis of  $\cal{H}_A$,
and choose $U_B\in O(\cal{H_B})$ be the n'th root of the unity:
$U^n_{B}=I_{B}$. We can than construct the n-level stator
\beq
{\cal{S}}=
|0_A\ra\otimes I_{B} + |1_A\ra\otimes U_{B} + \cdots |n-1_A\ra \otimes U^{n-1}_{B}
\eeq
The relevant eigenoperator equation than becomes
\beq
V_A  {\cal{S}} = U_B  {\cal {S}}
\eeq
where  $V_A$ is a shift operator
defined by:
$ V_A|m_A\ra = |(m-1)_A\ra$, $m=1\cdots n-1$,  and
$V_A|0_A\ra = |(n-1)_A\ra$.
By operating with $V_A + V^\dagger_A$ we then obtain the hermitian
eigenoperator $U_B+U^\dagger_B$, and  acting with
$i(V_A-V_A^\dagger)$
yields the eigenoperator $i(U_B-U_B^\dagger)$.
Similarly we can construct any powers of $U_B+U_B^\dagger$ and $i(U_B-U_B^\dagger)$.

We can further generalize our construction to the case that Bob has
at hand several systems  (which may be removed from each other)
of arbitrary dimension. We discuss this case in section 6.

\section{ Preparation of Stators}

We have seen that stators allow us to obtain state-independent
relations between Alice's actions and their result on Bob's state.
We proceed then to describe the process that will be referred to as
 ``preparation''
of a stator.
Hence, given by an  unknown state, $|\Psi_B\ra \in
{\cal H}_B$, and some shared entangled state $|{\rm ent}\ra$,
our aim is
transform this initial state by performing some LOCC operation into
\beq
|{\rm ent}\ra \otimes|\Psi_B\ra \to {\cal S}|\Psi_B\ra
\eeq

We first describe in details the simplest case in which Alice
and Bob use one ebit of shared entanglement to prepare a 2-level stator
as depicted in Figure 1.
The initial state at the hands of Alice and Bob is in this case
\beq
{1\over\sqrt2}\biggl(|0_a0_b\ra + |1_a1_b\ra
\bigg)\otimes |\Psi_B\ra
\eeq
For  practical purposes, in the following
we will denote by the small letters,  $a$ and $b$, the
shared ancillary entangled systems of Alice and Bob respectively.

\begin{figure} \epsfxsize=4.5truein
      \centerline{\epsffile{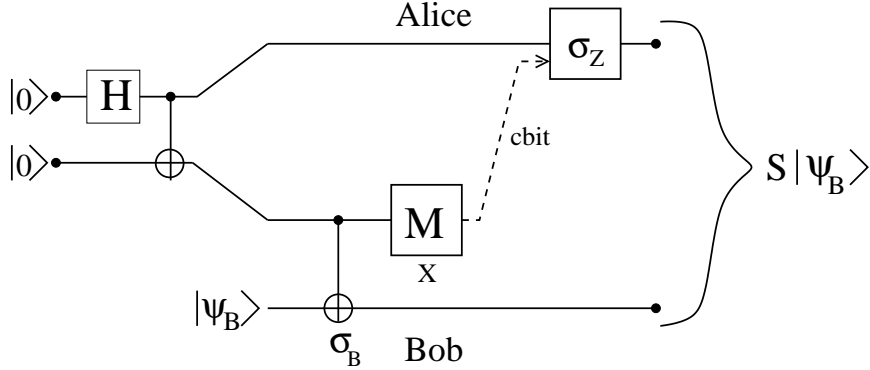}}
\vspace {0.5cm}
  \caption[]{ Preparation of a stator acting on Bob's state.}
    \label{prepare} \end{figure}

Bob starts by performing a CNOT interaction (with respect to $\sigma_{n_B}$)
between the qubit (b) and his state $|\Psi_B\ra$, described by the unitary transformation
\beq\label{cnot}
U_{bB} = |0_b\ra\la 0_b| \otimes I_{B} + |1_b\ra\la1_b|\otimes
\sigma_{n_B} \eeq
Here $\sigma_{n_B}$ is an operator acting in
${\cal H}_B$ satisfying $\sigma_{n_B}^2=I_{B}$. ( ${\cal H}_B$
need not be 2-dimensional; for instance Bob's system may contain
several spins,  in which case $\sigma_{n_B}=
\sigma_{n_B}^1\sigma_{n_B}^2 \cdots$).

This yields the state
\beq
{1\over\sqrt2}\biggl(|0_a0_b\ra\otimes I_{B}
 + |1_a1_b\ra \otimes\sigma_{n_B} \biggr)|\Psi_B\ra
\eeq
Next he performs a measurement of $\sigma_x$ of the
 entangled qubit to project out a certain value.   The resulting state is now
\beq\label{resx}
{1\over2}\bigl(|0_a\ra \pm |1_b\ra\bigr)\otimes \biggl(|0\ra_a\otimes I_B \pm |1\ra_a\otimes\sigma_{n_B} \biggr) |\Psi_B\ra
\eeq
Finally Bob informs Alice what was the result of his measurement
 by sending Alice one classical bit of information.
For the case that $\sigma_x=-1$ Alice performs
a trivial $\pi$ rotation around the $\hat z$ axis and flips the $-$ sign
to a $+$ sign. The resulting state of the system is now given by
\beq\label{Spsi}
{1\over2}\bigl(|0_a\ra \pm |1_b\ra\bigr)\otimes \biggl(|0\ra_a\otimes I_{B} + |1\ra_a\otimes\sigma_{n_B} \biggr) |\Psi_B\ra
\eeq

Since Bob's previously entangled qubit factors out,
the final state of Alice's qubit and Bob's system  can be obtained by
letting the stator
\beq
{\cal S}= |0_a\ra \otimes I_{B}+  |1_a\ra\otimes \sigma_{n_B}
\eeq
act  on $|\Psi_B\ra$.
This completes the preparation of a 2-level stator ${\cal S}$ which
now operates on Bob's system.

We further discuss preparation of n-level stators
in connection to remote operations on an n-level system in section 5.

\section{Remote Unitary transformations}

Suppose that Bob has a system in the unknown state $|\Psi_B\ra$
on which Alice and Bob wish to act on with a unitary transformation described
by a rotation
\beq
U_B= e^{i\alpha \sigma_{n_B}}
\eeq
with $\sigma_{n_B}^2=I_B$ and $\alpha$ a real arbitrary number.

We will now show that the transformation (20) can be performed, provided that
the generator $\sigma_{n_B}$ is known to Bob,
and the parameter  $\alpha$ of rotation is know to Alice.
To this end, they start by using a shared ebit to prepare,
 as described in the previous section,  the stator
\beq
{\cal S} = |0_a\ra \otimes I_B + |1_a\ra\otimes \sigma_{n_B}
\eeq
which operates on Bob's state. $\sigma_{n_B}$ enters here as
a result Bob's choice to perform a CNOT with respect to $\sigma_{n_B}$
as in eq. (\ref{cnot}).

Next, Alice performs on her qubit a unitary transformation
\beq
U_a= e^{i\alpha \sigma_{x_a}}
\eeq
where $\sigma_{x_a}|0_a\ra= |1_a\ra$ and $\sigma_{x_a}|1_a\ra= |0_a\ra$.
Using the fact that when acted with  $\sigma_{x_a}$ the stator
satisfies an eigenoperator equation with an eigenoperator
$\sigma_{n_B}$ we have
\beq
e^{i\alpha \sigma_{x_a}} {\cal S} =e^{i\alpha \sigma_{n_B}}  {\cal S}
\eeq
Hence after the rotation
the state is
\beq
\biggl( |0_a\ra \otimes I_B + |1_a\ra\otimes \sigma_{n_B}
\biggr) e^{i\alpha \sigma_{n_B}} |\psi_B\ra
\eeq
Depending upon the final state of Alice's qubit,
they managed to produced the required rotation,
modulo possible extra trivial rotations.
To eliminate these rotations, Alice measures the state of her qubit.
If it is $|0_a\ra$, we have produced the required transformation.
If it turns out to be in the state $|1_a\ra$ she needs to inform
Bob to perform a trivial $\pi$ rotation, $U_\pi=\exp(i\pi \sigma_{n_B}/2)$,
which corrects for the extra $\sigma_{n_B}$ above.
This completes the process.

\begin{figure} \epsfxsize=5truein
      \centerline{\epsffile{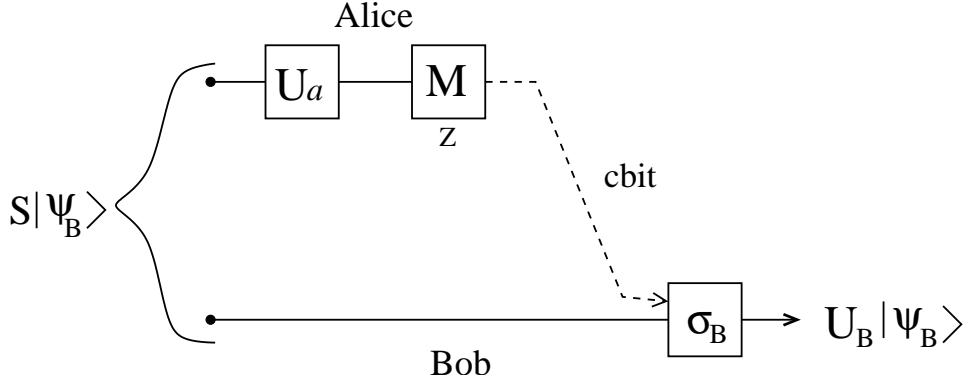}}
\vspace {0.5cm}
  \caption[]{ Usage of a stator to operate a remote rotation.}
    \label{rotate} \end{figure}

The resources that Alice an Bob require for remote rotation applied on a 2-level
system are hence, one e-bit of shared entanglement and two cbits.
They communicate
one cbit first from Bob to Alice to prepare the stator, and one cbit
from Alice back to Bob to complete the required rotation with
probability 1.  For both cbits we have that $p(1)=p(0)=1/2$, i.e. they are unbiased.
Therefore the exchanged classical communication contains no information on the state of
Bob or the angle of rotation.

The role of the exchanged cbits  is as follows:
the first cbit is needed in order to obtain the correct
stator (fix the sign in eq. (\ref{resx})). Without this one would have  obtained
with probability 1/2 the correct rotation $U$ and with probability
$1/2$ the rotation $U^\dagger$.
(For the case of remote measurements discussed in section 7. this
uncertainty in the sign may be irrelevant initially and
may be corrected at later stage of the process.) The second cbit sent from Alice to Bob
is clearly needed from causality requirement. A process that
uses less then one
cbit of communication from Alice to Bob clearly violates causality.

\section {Remote operations on n-level systems}

We  now apply our method for the case of an n-level system.
First we identify the n-level stator
with the  appropriate generator of rotations as an
eigenoperator. To prepare this stator, Alice and Bob apply
LOCC on their shared entangled state and Bob's n-level system.
Next Alice performs a unitary transformation on half of the entangled
pair on her side, followed by a measurement, and informs Bob via a
classical channel how to correct his system to complete the rotation.

As we shall shortly see the required resources
in this case are two maximally entangled n-level systems, and two
classical ``nits'' (each containing $n$ possible values),
one sent from Bob to Alice to complete the preparation, and the
second from Alice back to Bob to complete the remote operation.
However, unlike the 2-level case, the number of possible
unitary operation per given entangled n-level pair is here
larger and given by any general linear combinations of
$n-1$ generators. The rotation around a given axis
is one of the possible operations.

To illustrate this let us first demonstrate the process for the case
$n=3$ of a spin one particle.
For a rotation around the $z$- axis (with the axis of rotation
been chosen as  before by Bob) we need
to identify a stator that satisfies the eigenoperator equation

\beq
A  {\cal S} = L_{Z}{\cal S}
\label{lz}
\eeq
where $L_Z$ is the appropriate generator of rotation.
When applying $A^2$ on the stator we get
$A^2{\cal S} = L_{Z}^2{\cal S}$. Therefore  $L_Z^2$ is
another eigenoperator of ${\cal S}$. Since $L_Z^3=L_Z$ these are the only
eigenoperators.

Since for $n=3$ we have two distinct eigenoperators,
the most general remote transformation which we are able to construct, using
two maximally entangled 3-level systems (qutrit), has the form
\beq
U_B= e^{i(\alpha L_Z + \beta L_Z^2)}
\eeq
where $\alpha$ and $\beta$ are chosen by Alice.

Recalling the discussion in
section 2. the appropriate ${\cal S}$ for this case is a 3-level stator
of the form
\beq\label{3ls}
S=|0_a\rangle\otimes{I_{\Psi_B}}+|1_a\rangle\otimes{U_{\Psi_B}}
+|2_a\rangle\otimes{U^{2}_{\Psi_B}}
\label{3stator}
\eeq

where the requirement $U_{\psi_B}^3=I_{\Psi_B}$ dictates the form
\beq
U_{\Psi_B}= e^{{2\pi i\over 3} L_z}
\eeq
(Here we used the subscript $\Psi_B$ in $U_{\Psi_B}$ in order to distinguish between the full
remote operation $U_B$ applied by Alice and Bob and local transformations
$U_{\Psi_B}$ applied by Bob).
Since for a spin one particle we have $e^{i\theta L_z} =
1 + i\sin \theta L_z + L_z^2(cos\theta - 1)$,
we identify the operator $A$ in eq. (25) as
\beq
A = {1\over 2i\sin{2\pi\over 3} } \bigl(V - V^\dagger)
\eeq
where $V$ and $V^\dagger$ are the raising and lowering operators defined
in section 2.

Having identified the required stator and the operators $A$, we next
describe the preparation and rotation process.
We begin with a shared pair of maximally entangled qutrits and Bob's state
$|\Psi_B\ra$:
\beq
\biggl(|0_a0_b\ra+|1_a1_b\ra+|2_a2_b\ra\biggr) |\Psi_B\ra
\eeq
Bob applies the unitary operation $U_{bB}$ on his state
and his half (b) of the entangled pair:
\beq
U_{bB}=|0_b\ra\la0_b|\otimes I_{\Psi_B} + |1_b\ra\la1_b|\otimes U_{\Psi_B}
 + |2_b\ra\la 2_b|\otimes U_{\Psi_B}^2
\eeq

This results with the state
\begin{eqnarray}\label{bS}
|\Psi_{tot}\rangle=[|0_a0_b\rangle\otimes{I}_{\Psi_B}+|1_a1_b\rangle
\otimes U_{\Psi_B}+|2_a2_b\rangle\otimes{U_{\Psi_B}^2}]|\Psi_B\rangle
\end{eqnarray}
Now to generate the stator (\ref{3stator}),
we need to eliminate Bob's entangled particle.
Hence Bob measures his particle $b$ in the following basis:
\begin{eqnarray}
|0'_b\rangle=\frac{1}{\sqrt{3}}(|0_b\rangle+|1_b\rangle+|2_b\rangle)~~~~~~~~~\nonumber\\
|1'_b\rangle=\frac{1}{\sqrt{3}}(|0_b\rangle+e^{\frac{2\pi i}{3}}|1_b\rangle+e^{\frac{4\pi i}{3}}|2_b\rangle)\\
|2'_b\rangle=\frac{1}{\sqrt{3}}(|0_b\rangle+e^{\frac{4\pi i}{3}}|1_b\rangle+e^{\frac{2\pi i}{3}}|2_b\rangle)\nonumber
\end{eqnarray}
Let's rewrite the state (\ref{bS}) in the terms of the new basis
vectors:

\begin{quote}
\centering

\begin{eqnarray}
|\Psi_{tot}\rangle=\{~~~~~
[|0_a\rangle\otimes { I_{\Psi_B}}+~~~~|1_a\rangle\otimes
{U_{\Psi_B}}+~~~~|2_a\rangle\otimes{U^{2}_{\Psi_B}}]
|0'_b\rangle~~~~~~~~~~\nonumber\\
~~~~~~~+~~
[|0_a\rangle\otimes{I_{\Psi_B}}+e^{\frac{2\pi i}{3}}|1_a\rangle\otimes{U_{\Psi_B}}+e^{\frac{4\pi i}{3}}|2_a\rangle\otimes{U^{2}_{\Psi_B}}]|1'_b\rangle~~~~~~~~~~\\
~~~~~~~+~~
[|0_a\rangle\otimes{I_{\Psi_B}}+e^{\frac{4\pi i}{3}}|1_a\rangle\otimes{U_{\Psi_B}}+e^{\frac{2\pi i}{3}}|2_a\rangle\otimes{U^{2}_{\Psi_B}}]|2'_b\rangle~~\}|\Psi_B\rangle
\nonumber
\end{eqnarray}

\end{quote}
\noindent
According to one of the three particular outcomes of Bob's
measurement of the particle $b$ the state of Alice's particle $a$
and Bob's particle $\Psi_B$ evolves to
\beq
\label{state}
[|0_a\rangle\otimes {I_{\Psi_B}}+|1_a\rangle\otimes{U_{\Psi_B}}
+|2_a\rangle\otimes{U^{2}_{\Psi_B}}]|\Psi_B\rangle
\eeq
or to the states
$$[|0_a\rangle\otimes I_{\Psi_B}+e^{\frac{2\pi i}{3}}|1_a\rangle
\otimes{U_{\Psi_B}}+e^{\frac{4\pi i}{3}}|2_a\rangle\otimes{U^{2}_{\Psi_B}}]|\Psi_B\rangle$$
and
$$[|0_a\rangle \otimes{I_{\Psi_B}}+e^{\frac{4\pi i}{3}}|1_\rangle\otimes{U_{\Psi_B}}+e^{\frac{2\pi i}{3}}|2_a\rangle\otimes{U^{2}_{\Psi_B}}]|\Psi_B\rangle.$$
Bob transmits this classical outcome (classical "trit") to Alice.
Notice that the three results appear with equal probability of $1/3$
hence the classical trit is unbiased.
In the last two cases Alice performs the following transformations
on her particle $a$ in order to correct the state to the form
(\ref{state}):
$C_{1}=|0_a\rangle\langle{0_a}|+e^{\frac{4\pi i}{3}}|1_a\rangle\langle{1_a}|+e^{\frac{2\pi i}{3}}|2_a\rangle\langle{2_a}|$
and
$C_{2}=|0_a\rangle\langle{0_a}|+e^{\frac{2\pi i}{3}}|1_a\rangle\langle{1_a}|+e^{\frac{4\pi i}{3}}|2_a\rangle\langle{2_a}|$
respectively. We can interpret (\ref{state}) as the stator (\ref{3ls})
operating on the state $|\Psi_B\rangle$.
This completes the preparation process.

In order to generate a general rotation, Alice acts on here particle with the
unitary operator
\beq
U_a= e^{i(\alpha A + \beta A^2)}
\eeq
and performs a measurement to  collapses the
state into one of the states $|n_a\ra$. Notice that as in the preparation
process the results are again unbiased. She then sends a classical trit to
and informs Bob the result of her  measurement.
For the cases that Alice obtained  $|1_a\ra$ or  $|2_a\ra$, Bob then
performs the rotations $U_B^2$ and $U_B$, respectively.
This completes the procedure or generating a remote rotation.

The above procedure can be applied for an arbitrary $n$-level
system. The maximally entangled state of two qutrits is then replaced
by a maximally entangled pair of n-level systems. After applying the
interaction $U_{bB}$ the total state becomes
\begin{equation}\label{fs1}
|\Psi_{tot}\rangle=[\sum_{m=0}^{n-1}|n_an_b\rangle\otimes{U^m_{\Psi_B}}]
|\Psi_B\rangle
\end{equation}
with
\beq
U_{\Psi_B} = e^{{2\pi i\over n} L_Z}
\eeq
and $L_Z$ the appropriate rotation generator for the n-level system.

Bob then performs a measurement
of his half of the entangled pair $b$ in the following basis:

\beq
|m_b'\rangle=\frac{1}{\sqrt{n}}
\sum_{m_b=0}^{n-1}e^{\frac{2\pi i}{n}m'_bm_b}|m_b\rangle
\eeq
where $m'_b=0...n-1$.
He sends to Alice one ``nit'' to inform her which of the n-possible
outcomes was obtained.
Alice on her side operates the relevant unitary operation.
It can be shown that for $n>3$ the relevant operator $A$
in equation (\ref{lz}) becomes a linear combinations
of powers of $V-V^\dagger$ for $n$ odd,  and of $V+V^\dagger$ for
even $n$. The total number of independent combinations is $n-1$.
To complete the process Alice then performs a measurement and
send Bob one classical nit.
This enables him to perform one of the
operations $U_{\Psi_B}^m$, $m=0...n-1$
which complete the process.

To summarize: for an n-level system, we use the resources of one  pair of
maximally  entangled  n-level system and a two way classical communication
of one nit in each direction. This enables to apply a general remote
transformation of the form
\beq
U_B = e^{i(\alpha_1 L_Z + \alpha_2 L_Z^2 +.... \alpha_{n-1} L_Z^{n-1})}
\eeq
where Bob determines the axis $Z$, and Alice determines the $n-1$ angles $\alpha_i$.

\section{Operations on multiple systems}

Consider next the case that on Bob's side we have $N$ distinguishable
separate systems in some arbitrary state $|\Psi_{B_{1\cdots N}}\ra$:
\beq
|\Psi_{B_{1\cdots N}}\ra \in {\cal H}_{B_1}\otimes \cdots \otimes {\cal H}_{B_N}
\eeq
with ${\rm dim}{\cal H}_{B_i} = n_i$.
The $N$ systems may be distributed to $N$ different
remote spatially separated locations denoted by B$_i$.

To examine the operations possible by our method we
 further assume that we distribute between Alice and B$_i$ $N$
maximally entangled pairs as depicted in depicted in figure 3.
For a given system of dimensionality $n_i$ we match
a maximally entangled $n_i$-level pair shared between Alice and B$_i$.

\begin{figure} \epsfxsize=3.5truein
      \centerline{\epsffile{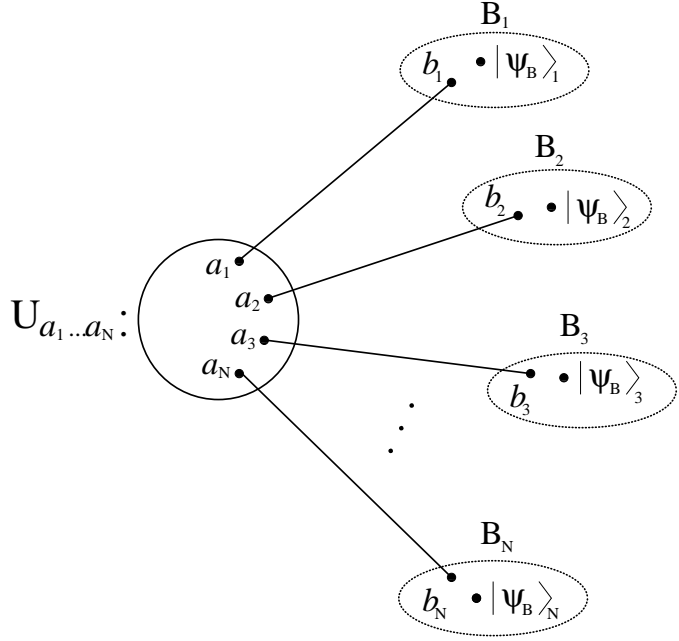}}
\vspace {0.5cm}
  \caption[]{ Remote operation on $N$ distributed systems.}
\label{rotate2} \end{figure}

Clearly we now can repeat our method and generate $n_i-1$
operations on the $i$'th system by using the shared entangled
pairs to prepare $N$ stators, each one connecting
between Alice and the system B$_i$.
 However it now turns out that with $N$
stators at hand we can generate an exponentially
larger class of operations, most of them corresponding to
interactions between several remote subsystems.

To exemplify this, consider first the simplest case of $N$ 2-level
(spin-half) systems.
In this case the resources needed are $N$ shared ebits between Alice
and B$_i$ and classical bidirectional communication of
$2N$ classical bits: two cbits between Alice and a given B$_i$.
As before each B$_i$ has the choice of fixing the local axis of
rotation which fixes $N$ generators
  $\sigma_{n_{B_i}}$, $i=1...N$.

We can repeat the preparation of a stator ${\cal S}_i$
for each spin separately as described in section 3.
The total stator is then
\beq
{\cal S}_{tot} = \otimes_{i=1}^{N} \biggl( |0_{a_i}\ra \otimes
I_{\Psi_{B_i}}  +
 |1_{a_i}\ra\otimes \sigma_{n_{B_i}} \biggr)
\eeq

The above stator satisfies an eigenoperator equations
\beq
\sigma_{x_i}{\cal S}_{tot} =  \sigma_{{B_i}}{\cal S}_{tot}
\eeq

However since the different $N$ generators commute, we also
have that  any {\em product}
 of separate eigenoperators is also an eigenoperator.
The total number of eigenoperators is then
\beq
\sum_{m=1}^N C^m_N   = 2^N -1
\label{num1}
\eeq
It follows then that Alice has the freedom of selecting the $2^N-1$
angles that generate rotations and interactions between the spins.

For example, the most general remote operation for the case $N=3$
becomes
\beq
U_B = \exp\biggl[ i\sum_{m=1}^3 \alpha_m\sigma_{B_m} +
i{1\over2}\sum_{m \ne n} \beta_{mn} \sigma_{B_m}\sigma_{B_n}
+i \gamma \sigma_{B_1}\sigma_{B_2}\sigma_{B_3} \biggr]
\eeq

We can easily apply our method for any configuration of  $N$ separated
$n_i$-levels systems. (In general $n_i$ may not be equal.)
Let us consider the case with $n_i=n$ for all $i$.
Then the total number of operators is easily computed to be
\beq
\sum_{m=1}^N(n-1)^m C^m_N  = n^N -1
\label{num2}
\eeq
 Therefore with the aid of
$N$ pairs of $n$-level maximally entangled pairs and bidirectional
classical communication of $2N$ nits we can apply
$n^N-1$ remote operations.

Finally, we note that the $N$ separated subsystems can be viewed as
a single system of dimensionality $D = \otimes_{i=1}^N n_i$.
Hence by the results
of the previous section, we can use one D-level stator
to act on the system as a whole. The number of distinct operation will
then given by $D-1$, in agreement with the results obtained
in eqs. (\ref{num1},\ref{num2}).

\section{Generating remote interactions and measurements}

In the last section we have already seen examples where Alice can act remotely
on several spatially separated systems and effectively generate
an interaction between remote subsystems.
For instance for two remote spins systems
Alice can use two ebits and four cbits to generate
the interaction
\beq
U_{B_1,B_2} = e^{i\alpha \sigma_{B_1}\sigma_{B_2}}
\eeq
Here the local axes of rotation, $\bar n_i$ ($\sigma_{B_i} =
\bar n_i \cdot \bar \sigma_{B_i}$), are determined locally by the local
observers B$_i$, and the coupling strength $\alpha$ is controlled by
Alice.

There is yet another simple method to generate remote interaction
between Bob's system and a system $A$ located with Alice.
Inspecting  eq. (10), we note that in fact the
angle $\alpha$ can be promoted  to an operator acting
on a system $A$ of Alice. Hence in the case of a 2-level stator,
with an eigenoperator $\sigma_{n_B}$ we have also the relation
\beq
e^{i\lambda O_A \sigma_{x_A}} {\cal S} = e^{i\lambda O_A \sigma_{n_B}}
{\cal S}
\label{int}
\eeq
where the stator ${\cal S}$ is defined as in eq. (7),
and $O_A$ is an hermitian operator acting on an arbitrary dimensional
   system $A$ of Alice.

A simple generalization of  the procedures in sections 3. and 4.
now allows performing remote interaction between separate systems.
The only modification needed is to replace the unitary rotation
performed by Alice to her half of the entangled pair, ({\it a}), with the
unitary operation
\beq
U_{Aa} =e^{i\lambda O_A \sigma_{x_a}}
\label{int2}
\eeq
acting on ({\it a}) and on her system A. (For the n-level case
$\sigma_{x_a}$ needs to be replaced by an appropriate operator, e.g.
the operator $A$ defined in eq. (29)).

For example suppose Alice's system is another spin-half particle,
and we wish to apply remotely a CNOT operation \cite{resources}
between Alice's and Bob's spins.
To this end we need to apply the transformation
\beq
U_{AB} = e^{{i\pi\over 4} \sigma_{x_A}(1-\sigma_{x_B})}
\eeq
But this is the special case of applying the transformation
(\ref{int}) while taking  $\lambda=\pi/4$, $O_A=\sigma_{x_A}$
 and  $n_B=x_B$, followed with a simple local rotation
$e^{{i\pi\over 4} \sigma_{x_A}}$.

As a special case of remote interactions we can further consider
remote measurements. Hence Alice's system A will be considered
as a measuring device . We can use another spin as a measuring device
(pointer) or let us introduce a continuous measuring
device with conjugate coordinates $P$ and $Q$, where $P$ plays
the role of the ``pointer''.

Let us describe a remote ``Stern-Gerlach'' measurement
of Bob's spin system
along a certain direction. Alice informs  Bob to fix the axis
$n_B$ according to the direction she wishes to perform the
measurement. After completing the preparation of the stator
she applies the unitary operation
\beq
U_{{\rm MD},a} = e^{iQ\sigma_{x_a}}
\eeq
that yields the state
\beq
{\cal S}e^{iQ\sigma_{n_B}} |\Psi_B\ra| MD\ra
\eeq
She can now observe the variable $P$ of the measuring device
and read the outcome of the measurement.
The final state of the system $B$ still does not correspond to the
outcome of the measurement. For this she needs to measure
the entangled particle $({\it a})$ and communicate to Bob the result
(not the result of the measurement!). Bob uses this random information
to corrects the
state to match the result of the measurement.

We conclude with several comments. The remote
measurement process can be in fact completed instantaneously on a space-like surface;
 Alice does not need to wait
to obtain a classical bit to perform her measurement.
In this case  Alice generates the operation
$\exp{(\pm i Q \sigma_{n_B})}$
with probability 1/2 for each $\pm$ possibility. Hence, in accordance with causality,
     the result of the measurement
can be interpreted only {\em after} she obtains the classical bit
from Bob. This approach can be easily generalized for
general systems as well as for performing measurements
non-local observables.

Finally, it is interesting to note that the present method consumes less
entanglement resources (one ebit instead of two, and two cbits
instead of four) compared with methods using teleportations.
It can also be shown that some non-local measurements can be
performed using the present method but cannot by using teleportation.

\section{Conclusion}
We presented a systematic method for constructing
deterministic remote operations on single and multiple systems
of arbitrary dimensions.
Our approach requires bidirectional  classical communication of unbiased bits
between the parties
and leaves the control over the generators that act on each
system at the hands of the local observers.
In this way the control over the full structure of
the unitary operation is split between several remote observers.
It is also worth to mention that when the local
information is kept secrete,
the operations cannot be achieved using
teleportation-like schemes. These properties may be helpful for constructing new cryptographic
tools.

To facilitate the construction of remote operations we have
introduced a new object -- the  stator -- which describes
correlations between states of one system and operations acting
on an arbitrary state of another remote system.
We hope that stators may turn out useful  for other problems
regarding the relation between entanglement and
remote interactions.

\vspace {.3cm}
B. R. would like to thank M. Plenio and S. Huelga for discussions.
We acknowledge the Support from grant 471/98 of the Israel Science Foundation, established by the Israel Academy of Sciences and Humanities.

\vspace{.3cm}
Note: After this work has been completed, we have learned of
other results obtained independently by S. Huelga, M.B. Plenio and
J.V. Vaccaro quant-ph/0107110,
and by Chui-Ping Yang and J. Jea-Banacloche quant-ph/0107100.


\end{document}